\documentclass[a4paper, amsfonts, amssymb, amsmath, reprint, showkeys, nofootinbib, twoside,superscriptaddress,citeautoscript,citeautoscript]{revtex4-1}
\usepackage[english]{babel}
\usepackage[utf8]{inputenc}
\usepackage{mathtools}

\usepackage[colorinlistoftodos, color=green!40, prependcaption]{todonotes}
\usepackage{amsthm}
\usepackage{mathtools}
\usepackage{physics}
\usepackage{xcolor}
\usepackage{graphicx}
\usepackage[left=16mm,right=16mm,top=20mm,bottom=22mm,columnsep=15pt]{geometry} 
\usepackage{adjustbox}
\usepackage{placeins}
\usepackage[T1]{fontenc}
\usepackage{lipsum}
\usepackage{csquotes}

\usepackage[pdftex, pdftitle={Article}, pdfauthor={Author}]{hyperref} 
\usepackage{float}
\newcommand{\kB}{k_{\mathrm{B}}}
 
\setcitestyle{super}
\def\kbt{k_{\mathrm{B}}T}

\newcommand{\comment}[1]{}
\newcommand{\revision}{\textcolor{black} } 
\newcommand{\revisionn}{\textcolor{black} } 

\begin{document}
\title{Splitting probabilities as optimal controllers of rare reactive events}
\author{Aditya N. Singh}
\email[Electronic mail:]{ansingh@berkeley.edu}
\affiliation{Department of Chemistry, University of California, Berkeley, CA 94720, USA}
\affiliation{Chemical Sciences Division, Lawrence Berkeley National Laboratory, Berkeley, CA 94720, USA}

\author{David T. Limmer}
\email[Electronic mail:]{dlimmer@berkeley.edu}
\affiliation{Department of Chemistry, University of California, Berkeley, CA 94720, USA}
\affiliation{Chemical Sciences Division, Lawrence Berkeley National Laboratory, Berkeley, CA 94720, USA}
\affiliation{Materials Science Division, Lawrence Berkeley National Laboratory, Berkeley, CA 94720, USA}
\affiliation{Kavli Energy Nanoscience Institute at Berkeley, Berkeley, CA 94720, USA}

\date{\today} 

\begin{abstract}
The committor constitutes the primary quantity of interest within chemical kinetics as it is understood to encode the ideal reaction coordinate for a rare reactive event. We show the generative utility of the committor, in that it can be used explicitly to produce a reactive trajectory ensemble that exhibits numerically exact statistics as that of the original transition path ensemble. This is done by relating a time-dependent analogue of the committor that solves a generalized bridge problem, to the splitting probability that solves a boundary value problem under a bistable assumption. By invoking stochastic optimal control and spectral theory, we derive a general form for the optimal controller of a bridge process that connects two metastable states expressed in terms of the splitting probability. This formalism offers an alternative perspective into the role of the committor and its gradients, in that they encode forcefields that guarantee reactivity, generating trajectories that are statistically identical to the way that a system would react autonomously.  
\end{abstract}
\maketitle


The existence of a reaction coordinate, a function of a system's dynamic variables that captures the intricacies of a molecular transformation, was a guiding principle for development modern rate theory.\cite{kramers1940brownian,eyring1935activated} Through advances in statistical physics, an ideal reaction coordinate has been formally defined as the committor or splitting probability\cite{bolhuis2002transition,vanden2006towards}, and through a range of numerical methods\cite{peters2006obtaining,ma2005automatic,khoo2019solving,rotskoff2022active,hasyim2022supervised,coifman2008diffusion,thiede2019galerkin,escobedo2009transition,yuan2024optimal,gao2023transition}, it has become tractable to estimate it without prior physical insight into the reaction. Not only does knowledge of the committor provide mechanistic insight into reactions, but it can be used to evaluate their rates. As reactive events often represent the slowest relaxational processes of a molecular system, it is common wisdom that the committor provides an optimal means of sampling complex systems configurational spaces. Here, using advances in the theory of stochastic optimal control, we prove this conventional wisdom, and show that the commitor is an ideal importance sampler for reactive processes. In particular, the gradient of its logarithm can be used to generate an ensemble of driven trajectories that is statistical indistinguishable from the ensemble of trajectories conditioned to react. 
%

The committor is a conditional probability, a nonlinear function of the whole phase space rendering it difficult to interpret for complex systems. From the developments of Transition Path Theory,\cite{vanden2010transition} it has been grounded rigorously as the solution of specific partial differential equation, and much previous work has offered quantitative assessments of its properties.\cite{berezhkovskii2005one,peters2006obtaining,rhee2005one,peters2017reaction,louwerse2022information}
As an example, the gradients of the committor generate a vector field that point parallel along the reaction coordinate. \cite{berezhkovskii2005one,he2022committor} Since the dynamics underlying chemical systems are Markovian, an interpretation of the committor can be constructed within the application of \textit{Doob's h-transform}\cite{doob1984classical} that states that any Markov process under some conditioning can be decomposed into the unconditioned process with an additional external drift. The application of this theorem within the context of large deviation theory has been illustrated\cite{jack2015effective,chetrite2015nonequilibrium,chetrite2015variational} and employed in sampling rare fluctuations within steady-states.\cite{ray2018exact,das2019variational,nemoto2016population,nemoto2017finite,rose2021reinforcement,yan2022learning,das2021reinforcement,hartmann2013characterization} The external drift from
 Doob's h-transform is given by the solution of a stochastic optimal control problem, where it is the function that maximizes the overlap between the ensemble of trajectories of the conditioned process and those from the process evolving under the external drift, or the driven process.\cite{ray2018importance} For a reactive event, the conditioning is simply that the system starts in a reactant state and ends in a product state at some later but finite time, a form of conditioning known as generalized bridge problem.\cite{majumdar2015effective}

In this work, we show that when the reactant and product states are metastable, the optimal controller can be simply related to the committor. 
\revisionn{Formally, this work extends previous results on relations between splitting probabilities and optimal control of time-independent reactive trajectories in systems with auxiliary boundary conditions\cite{hartmann2012efficient,yuan2024optimal,zhang2014applications,lu2015reactive} to finite-time reactive trajectories that are defined throughout the whole state space without any boundary conditions.}
For the two systems considered, a low dimension exactly solvable model and the dissolution of a colloidal cluster, the controller is shown to generate trajectories that are nearly indistinguishable from that of the original reactive ensemble. This result provides an interpretation of the committor as the exponential of a potential, the gradients of which explicitly encode forces corresponding to the fluctuations of the system when it reacts naturally in finite time. Conceptually, this connection provides a mathematical basis for the long observed results of why biasing along the reaction coordinate accelerates kinetics. Practically, this work offers a way to solve the path sampling problem by solving the configuration sampling problem, since all the quantities required to construct the controller can be obtained from a range of state-of-the-art methods that only require access to the stationary distribution.

\section{Theory}

For simplicity we assume that the system is comprised of $N$ degrees of freedom, denoted $\mathbf{r}$, and evolves according to an overdamped Langevin equation
\begin{equation}\label{eq:1}
    \gamma_i  \dot r_i = F_i(\mathbf{r}) + \eta_i
\end{equation}
where $r_i$ is the position of the $i$'th degree of freedom, $F_i(\mathbf{r})$ is the drift acting on it, which we assume is given by the gradient of a confined potential $F_i(\mathbf{r}) = -\partial_{r_i} V(\mathbf{r})$, and $\gamma_i$ is the associated friction coefficient. The random force, $\eta_i$ is given by a Gaussian white noise with mean $0$ and variance $\langle \eta_i(0)\eta_j(t) \rangle = 2 \kbt \gamma \delta(t) \delta_{ij}$, where $\kbt$ is temperature times Boltzmann's constant. Working within the overdamped Langevin equation assumes a large friction limit, and can be relaxed with perturbation theory.\cite{singh2023variational,hartmann2014optimal} Generalizations to other Markovian dynamics are also possible.

\subsection{Forward and backward transition densities}
The evolution for the probability density generated by Eq.~\ref{eq:1} is given by the Fokker-Planck equation (FPE),\cite{zwanzig2001nonequilibrium}
\begin{align}\label{eq:2}
    \frac{\partial}{\partial t} &P(\mathbf{r},t) = \mathcal{L_{\mathbf{r}}} P(\mathbf{r},t)  \\
    &= \sum_{i}^N -\frac{1}{\gamma_i} \frac{\partial}{\partial r_i} F_i(\mathbf{r}) P(\mathbf{r},t) + \frac{\kbt}{ \gamma_i } \frac{\partial^2}{\partial r_i^2} P(\mathbf{r},t) \notag
\end{align}
where $\mathcal{L_{\mathbf{r}}}$ is the Fokker-Planck generator, and $P(\mathbf{r},t)$ is the probability of the system to be in position $\mathbf{r}$ at time $t$. Since this equation holds for any initial condition, it follows that the conditional probability, $P(\mathbf{r},t|\mathbf{r'},0)$, of starting at $\mathbf{r'}$ at time $t=0$, and ending at $\mathbf{r}$ at time $t>0$, also satisfies the Fokker-Planck equation,
\begin{equation}
 \frac{\partial}{\partial t} P(\mathbf{r},t|\mathbf{r'},0) = \mathcal{L_{\mathbf{r}}} P(\mathbf{r},t|\mathbf{r'},0)
\end{equation}
with the same generator. \revision{The conditional probability, $ Q( \mathbf{r}',t'| \mathbf{r},t )$, of being at some position $\mathbf{r}$ at some time $t$, given the terminal conditioning of the system to be at $\mathbf{r}'$ at a later time $t'>t$, satisfies a distinct though closely related equation, known as the backward Kolmogorov equation (BKE),}\cite{van1992stochastic}
\begin{align}\label{eq:3}
\frac{\partial}{\partial t} Q( \mathbf{r}',t' | \mathbf{r},t) = -\mathcal{\mathcal{L_{\mathbf{r}}^\dag}} Q(\mathbf{r}',t' | \mathbf{r},t) 
\end{align}
where $\mathcal{\mathcal{L_{\mathbf{r}}^\dag}}$ is equal to,
\begin{equation}
 \mathcal{\mathcal{L_\mathbf{r}^\dag}}=  \sum_{i}^N \frac{F_i (\mathbf{r})}{\gamma_i}   \frac{\partial}{\partial r_i}  + \frac{\kbt}{ \gamma_i } \frac{\partial^2}{\partial r_i^2} 
\end{equation}
the adjoint of the Fokker-Planck generator.

As linear equations, both the FPE and the BKE can be solved using an eigenvector expansion. However, as $\mathcal{L_{\mathbf{r}}}$ is not Hermitian, its eigenvectors and those for its adjoint are different, though they share eigenvalues. The Fokker-Planck generator satisfies an eigenvalue equation
\begin{equation}
\mathcal{L_{\mathbf{r}}} \psi^R_n(\mathbf{r}) = -\mu_n \psi^R_n(\mathbf{r})
\end{equation}
where $\mu_n$ is the $n$'th eigenvalue and $\psi_n^R(\mathbf{r})$ is the right eigenvector, while its adjoint satisfies 
\begin{equation}
\mathcal{L_{\mathbf{r}}^\dag} \psi^L_n(\mathbf{r}) = -\mu_n \psi^L_n(\mathbf{r})
\end{equation}
with $\psi^L_n(\mathbf{r})$ known as the left eigenvector.  The left and right eigenvectors can be related through an orthogonality condition,
\begin{equation}\label{eq:11}
    \int d\mathbf{r} \; \psi^L_n(\mathbf{r}) \psi^R_m(\mathbf{r}) = \delta_{nm}
\end{equation}
where $\delta_{nm}$ is a Kronecker delta function. From the Perron-Frobenius theorem, the first eigenvalue $\mu_1=0$, and the first right eigenvector is given by the steady-state density of the system $\bar{p}(\mathbf{r})$, which is the Gibbs-Boltzmann distribution if the force is given by the gradient of a conservative potential. It follows that 
\begin{align}
    \psi^R_1(\mathbf{r}) &= \bar{p}(\mathbf{r}) \qquad 
    \psi^L_1(\mathbf{r}) = 1
\end{align}
the dominate left eigenvector  by orthonormalization must be 1. This admits a normalization relation for the eigenvectors,
\begin{equation}\label{eq:10}
     \psi^L_n(\mathbf{r}) = \frac{\psi^R_n(\mathbf{r})}{ \psi^R_1(\mathbf{r})}
\end{equation}
relating the left and right to the stationary distribution. 
The solution to the BKE can computed by performing a spectral decomposition\cite{ryter1987eigenfunctions,coifman2008diffusion,roux2022transition},
\begin{equation}\label{eq:9}
 Q (\mathbf{r}',t_f | \mathbf{r},t) =   \sum_{n=1} \psi^L_n(\mathbf{r'}) \psi^R_n(\mathbf{r}) e^{-\mu_n (t_f-t)}
\end{equation}
while an analogous expression holds for the FPE. Thus given the left and right eigenvectors of $\mathcal{L_{\mathbf{r}}}$ and its associated eigenvalues, forward and backward transition probability densities can be evaluated.

\subsection{Time-dependent and steady-state committors}
We assume that the system exhibits metastability, and for simplicity that there are two dominant states denoted $A$ and $B$ \revision{that are formally defined as disjoint subsets of the whole state space and identified by the indicator functions}
\begin{equation}\label{eq:12}
h_X[\mathbf{r}(t)] = \begin{cases} 1 \quad \mathbf{r}(t)\in X\\
    0 \quad \mathbf{r}(t)\notin X \end{cases}
\end{equation}
\revisionn{where $X$ denotes the subsets $A$ or $B$}. The steady state probability density, $\bar{p}_X$ in $X$, is given by
\begin{equation}\label{eq:13}
    \bar{p}_X =\int d\mathbf{r} \; \bar{p}(\mathbf{r}) h_X(\mathbf{r})
\end{equation}
and $\bar{p}_A+\bar{p}_B \approx 1$ for such bistable system. \revisionn{Without loss of generality, we can define $k_{AB}(t_f)$ as a time-dependent reactive rate from $A$ to $B$ defined as 
\begin{align}\label{eq:rate}
    k_{AB} (t_f) &= \frac{1}{t_f}\frac{\langle h_{A}(0) h_B(t_f)\rangle}{\langle h_A(0) \rangle } \notag \\
    &= \frac{1}{t_f}\langle h_B(t_f) \rangle_{A}
\end{align} 
where the brackets denote average within the steady state, the subscript $A$ a conditioned average, and $k_{BA}(t_f)$ will denote the reverse rate. Provided metastability and a separation of local relaxation times and barrier crossing, the second eigenvalue is related to the reactive rate as,\cite{shuler1959relaxation}
\begin{equation}\label{eq:mu2}
\revisionn{
    \mu_2 \approx k_{AB}(t_f)+ k_{BA}(t_f) }
\end{equation} 
where the approximation lies within the choice of observation time $t_f$. For a bistable system $\mu_3 \gg \mu_2 \gtrsim 0$, manifesting the fact that relaxation within a basin of attraction occurs more rapidly, with rate $\mu_3$ \revisionn{than} transitions between basins. Hence, the rate grows linearly in time when $t_f$ is chosen $\mu_3^{-1} < t_f  \ll \mu_2^{-1} $, and further multiple transitions or backward transitions are exponentially unlikley.\cite{chandler1978statistical} For the rest of the paper, we assume that this approximation holds, and absorb the time-dependence of the rate for brevity.}

Given the Markov property of the equation of motion, it is straightforward to construct the probability of finding the system at some position $\mathbf{r}$ and time $t$ given it started in basin $A$ and ends in basin $B$ at some final time $t_f$. The likelihoods of such configurations are members of a reactive path ensemble, whose probability density $\rho_{AB}$ is given by
\begin{equation}\label{eq:4}
    \rho_{AB} (\mathbf{r},t) = \frac{p_A(\mathbf{r},t) q_B(\mathbf{r},t_f-t)}{\bar{\rho}_{AB}(t_f)}
\end{equation}
which is a product of 
\begin{equation}
p_A(\mathbf{r},t) = \frac{1}{\bar p_A}\int d\mathbf{r}' P(\mathbf{r},t|\mathbf{r'},0) h_A(\mathbf{r}') \bar{p}(\mathbf{r}')
\end{equation}
the probability of being at $\mathbf{r}$ and $t$ given the system started in $A$ at steady-state, and
\begin{equation}
q_B(\mathbf{r},t_f-t) = \int d\mathbf{r}' Q(\mathbf{r}',t_f|\mathbf{r},t) h_B(\mathbf{r}')
\end{equation}
the probability of  being at $\mathbf{r}$ and $t$ given the system ends in $B$ at time $t_f$. The normalization constant, $\bar{\rho}_{AB}(t_f)$, is nothing but the probability of a transition, which 
is related to the conditional reactive rate from $A$ to $B$, $\bar{\rho}_{AB}(t_f) =  k_{AB}t_f$. 

The conditioning in $\rho_{AB}$ is an example of a generalised Brownian bridge, with equation of motion given by Doob's h-transform,
\begin{align}\label{eq:5}
     \frac{\partial}{\partial t} \rho_{AB}(\mathbf{r},t)& = \mathcal{L_{\mathbf{r}}}\rho_{AB}(\mathbf{r},t) \\+& \sum_i \frac{2\kbt}{\gamma_i} \frac{\partial}{\partial {r_i}} \left[\frac{\partial}{\partial {r_i}} \ln q_B(\mathbf{r},t_f-t)\right] \rho_{AB}(\mathbf{r},t)\notag
\end{align}
which is the Fokker-Planck equation for the original process with an additional conservative drift. This 
implies that reactive trajectories can be generated directly, provided an additional force
\begin{equation}\label{eq:6}
    \gamma_i \dot{r}_i = F_i (\mathbf{r}) + 2 \kbt \frac{\partial}{\partial {r_i}} \ln q_B(\mathbf{r},t_f-t) + \eta_{i}
\end{equation}
related to the so-called time-dependent committor,  $q_B(\mathbf{r},t_f-t)$. The direct application of this equation to generate reactive paths has been used by Orland and co-workers,\cite{majumdar2015effective,koehl2022sampling} and in Variational Path Sampling,\cite{das2022direct} where approximate forms of $q_B(\mathbf{r},t_f-t)$ are used with trajectory reweighting to generate an ensemble of reactive trajectories and infer reaction rates.  Additionally, this expression has been used in diffusion models, where $\ln q_B(\mathbf{r},t_f-t)$ is known as the score function.\cite{song2021maximum}

There are few cases where $q_B(\mathbf{r},t_f-t)$  can be deduced exactly. However, for bistable systems, we have found that it is directly related to the splitting probability defined within Transition Path Theory, used in the characterization of reactive events. This follows by truncating an eigenvector expansion for $q_B(\mathbf{r},t_f-t)$ at second order, 
\begin{align}\label{eq:21}
q_B(\mathbf{r},\tau) &\approx \int d\mathbf{r'} p(\mathbf{r'}) (1+\psi_2^L (\mathbf{r}) \psi_2^L(\mathbf{r'})e^{-\mu_2 \tau}) h_B(\mathbf{r'}) \notag \\
& = \bar{p}_B \left (1 +b\psi_2 ^L(\mathbf{r}) e^{-\mu_2 \tau} \right )
\end{align}
 where $b$ given from the integral over $\mathbf{r}'$. This approximate form is valid for systems that exhibit a strong separation of timescales and have a large gap in their spectrum between $\mu_2$ and $\mu_3$, for times $\tau=t_f-t$ larger \revisionn{than} $1/\mu_3$. Acting the BKE operator on $\psi_2 ^L(\mathbf{r})$ and invoking the spectral gap of the eigendecomposition,\cite{berezhkovskii2004ensemble}
\begin{equation}\label{eq:19}
\mathcal{L_{\mathbf{r}}^\dag} \psi_2 ^L(\mathbf{r})  = - \mu_2 \psi_2 ^L(\mathbf{r}) \approx 0
\end{equation}
which for a metastable system, is equivalent to the existence of a pseudo-state-steady. \revision{This can be understood by considering the evolution of probability density initiated within one of the stable wells. If the second eigenvalue was set to zero, the transient probability density would be simply given by the steady-state probability density within the metastable basin it was initiated in, and zero everywhere outside of it.}

Equation \ref{eq:19} is precisely the equation solved by the splitting probability, the probability that a configuration at $\mathbf{r}$ visits $B$ before $A$. Denoting the splitting probability, or \textit{steady-state} committor as $\bar q_{B} (\mathbf{r})$, it satisfies
\begin{align}\label{eq:8}
   \mathcal{L_{\mathbf{r}}^\dag} \bar q_B(\mathbf{r}) = 0
\end{align}
 with boundary conditions $\bar q_{B} (\mathbf{r})=1$ for $\mathbf{r} \in  B$ and $\bar q_{B} (\mathbf{r})=0$ for  $\mathbf{r} \in  A$.  Indeed, one can show that $\psi_2 ^L(\mathbf{r})$ is linearly related to $\bar q_B(\mathbf{r})$, \cite{ryter1987eigenfunctions,berezhkovskii2004ensemble,bovier2016metastability,roux2022transition,chen2023discovering}
\begin{equation}\label{eq:20}
\psi_2^L(\mathbf{r}) =(b-a)\bar q_B(\mathbf{r})  + a
\end{equation}
with values for $\psi_2^L(\mathbf{r})$ of $a$ and $b$ within well $A$ and $B$, respectively. The boundary value problem for the committor and the generalized eigenvalue problem are so simply related here only because the information on the boundaries of the metastable states are sufficient to determine the function within them because both functions are constant within those domains\cite{ryter1987eigenfunctions}. \revision{This relation only holds when $B$ is defined within a metastable basin}. In App. \ref{sec:a1}, we solve for $a$ and $b$ and prove that $\psi_2^L(\mathbf{r})$ is generally constant within a metastable basin. Hence by rescaling the left eigenfunction, we have an expression relating the time-dependent and time-independent commitors, 
\begin{equation}\label{eq:26}
    q_B(\mathbf{r},\tau) = \bar q_B (\mathbf{r}) e^{-\mu_2 \tau} + \bar{p}_B \left (1-e^{-\mu_2 \tau} \right )
\end{equation}
or equivalently, a demonstration that the drift that generates reactive trajectories in a manner statistically equivalent to the native reactive path ensemble, as in Eq.~\ref{eq:5}, follows the gradient of the committor. Formally, this result confirms the conventional wisdom that biasing along the committor is optimal for generating rare events with a sharp statement that when it is dressed by an appropriate time dependent function it acts as a potential to generate reactive trajectories. Practically, given the preponderance of methods to approximate $\bar q_B (\mathbf{r})$ in complex systems, this relationship provides a means of translating those methods into well-controlled generative models. 

\revision{This is the main result of this work that provided the committor,  $\bar q_B (\mathbf{r})$, finite-time reactive trajectories can be generated in proportion to their correct statistical weights.}
\revision{We note that while these results are similar to the earlier work on committor and optimal control dynamics\cite{hartmann2013characterization,hartmann2012efficient,zhang2014applications}, a key distinction is the treatment of reactions as a generalized bridge problem that is defined in the whole state space with a conditioning that trajectories react in finite time\cite{bolhuis2002transition}. As opposed to the Dirichlet exit problem, this treatment leads to a  different optimal control problem with distinct solutions, and one whose treatment has as a range of metrics for analyzing the accuracy of the controlled dynamics that we will now discuss.}

\subsection{Trajectory averages and variational principle}
In order to use the results above, we need to work within a representation amenable to molecular simulation. We consider an ensemble of paths, and use the fact that transition probabilities can be expressed as integrals over stochastic trajectories. The transition probability between $\mathbf{r}''(t)$ and $\mathbf{r'}(0)$, is given by an integral over paths \revision{$\mathbf{R}(t)=\{\mathbf{r}(t') : 0\leq t' \leq t \}$, }
\begin{equation}
P(\mathbf{r}'',t|\mathbf{r'},0)=  \int \mathcal{D}[\mathbf{R}(t)]  \delta[\mathbf{r}''-\mathbf{r}(t)]\delta[\mathbf{r}'-\mathbf{r}(0)]e^{-U_0[\mathbf{R}(t)] }
\end{equation}
constrained to start at $\mathbf{r'}(0)$ and end at $\mathbf{r}''(t)$, weighted by the negative exponential of
\begin{equation}
U_0[\mathbf{R}(t)] = \sum_i \frac{1}{4 \kB T \gamma_i} \int_0^t  dt' \left [\gamma_i \dot{r}_i - F_i(\mathbf{r}) \right ]^2
\end{equation}
known as the Onsager-Machlup action,\cite{onsager1953fluctuations,machlup1953fluctuations} here for the Ito process in Eq.~\ref{eq:1}. \revisionn{Expectation values can be computed as averages over stochastic paths. For example, the equation in \ref{eq:rate} denotes an average with path weights given by $U_0[\mathbf{R}(t)] $.}

The stochastic process in Eq.~\ref{eq:6} motivates the consideration of a dynamics under an additional general control force $\boldsymbol{\lambda}$, such that the equation of motion for the $i$'th degree of freedom is
\begin{equation}\label{eq:lameom}
    \gamma_i  \dot r_i = F_i(\mathbf{r}) + \lambda_i(\mathbf{r},t)+ \eta_i
\end{equation}
where $\lambda_i$ is the $i$'th component of $\boldsymbol{\lambda}$. Since the noise statistics of the two processes Eq. \ref{eq:1} and Eq.~\ref{eq:lameom} are the same, we can use trajectory or Girsanov reweighting to reexpress Eq.~\ref{eq:rate} for the rate as\cite{kuznets2021dissipation}
\begin{align}\label{eq:31}
   k_{AB}t_f &=  \left \langle e^{-\Delta U_{\boldsymbol{\lambda}}}\right \rangle_{B|A,\lambda} \langle h_{B}(t_f) \rangle_{A,\lambda} \notag  \\
   &=  \left [\left \langle e^{\Delta U_{\lambda}}\right \rangle_{B|A,0} \right ]^{-1}\langle h_{B}(t_f) \rangle_{A,\lambda}
\end{align}
where the subscript $\lambda$ denotes an average over trajectories generated with Eq.~\ref{eq:lameom}, while $B|A$ denotes a conditional average for trajectories that start in basin $A$ and end in basin $B$. The reweighting factor in the average is a difference in the Onsager-Machlup actions between an ensemble driven with a control force and without one, 
\begin{align}\label{eq:30}
\Delta U_{\lambda}[\mathbf{R}(t)] =  \sum_{i}^N \Delta U^i_\lambda 
\end{align}
where 
\begin{equation}\label{eq:decomp}
\Delta U^i_\lambda = \int_0^{t_f} dt  \frac{2\lambda_i(\mathbf{r},t)(\gamma_i \dot r_i - F_i(\mathbf{r})) -\lambda_i^2(\mathbf{r},t)  }{4 \gamma_i \kbt}
\end{equation}
is the change in action for each component. This quantity can be recognized as the  Radon-Nikodym derivative between the driven and the original process or alternatively on average is the Kullback-Leibler divergence between the two path distributions. On applying Jensen's inequality, one can deduce bounds for the rate\cite{kuznets2021dissipation}
\begin{equation}\label{eq:32}
- \langle \Delta U_{\lambda}\rangle_{B|A,\lambda} \leq \ln \frac{k_{AB} t_f }{\langle h_{B}(t_f) \rangle_{A,\lambda} } \leq -\langle \Delta U_{\lambda} \rangle_{B|A,0} 
\end{equation}
which forms a set of variational principles for optimizing $\boldsymbol{\lambda}$. The first with the averaging done in the driven ensemble can be caste as a reinforcement learning problem,\cite{das2021reinforcement} while the second with the averaging done in the conditioned reference ensemble is a fitting problem.\cite{singh2023variational}
Using the Hamilton-Jacobi-Bellman equation, one can show the inequalities are saturated when
\begin{align}\label{eq:27}
    \lambda^*_i(\mathbf{r},\tau) &= 2 \kbt \; \frac{\partial}{\partial r_i}  \ln q_B(\mathbf{r},\tau)   \\
    &\approx 2 \kbt \; \frac{\partial}{\partial r_i}  \ln \left [\bar{q}_B(\mathbf{r})e^{-\mu_2 \tau}+\bar{p}_B \left (1-e^{-\mu_2 \tau} \right ) \right ]\notag 
\end{align}
the force associated with Doob's h-transform, or the time-dependent committor, and $\tau=t_f-t$.\cite{chetrite2015variational,chetrite2015nonequilibrium,das2022direct} Hence, agreement between the rate and the average of the $-\Delta U_{\lambda}$ computed within the two different reactive ensembles offers a metric to validate the accuracy of an approximate control force. Further, when the bound is saturated, its decomposition in Eq.~\ref{eq:30} and ~\ref{eq:decomp} provides a way to resolve the typical contribution of each reactive mode and their couplings to the rare event, elucidating mechanism.\cite{singh2023variational}

\section{Numerical Illustration}
We now demonstrate how the effective control force that we derived in terms of the splitting probability in Eq. ~\ref{eq:26} offers an accurate solution to the generalized bridge problem within the context of reactive path sampling. We will consider both an exactly solvable problem in one dimension, as well as an interacting system where the committor can be solved numerically exactly using a neural network ansatz. Equation \ref{eq:32} provides a simple metric for the accuracy of the control force, though we have also found it useful to consider the distribution of $\Delta U_\lambda$, $ P_\lambda(-\Delta U_{\lambda}) = \langle \delta [- \Delta U_\lambda + \Delta U_{\lambda}(\mathbf{R})]\rangle_{B|A,\lambda}$, generated with and without the control force, $\lambda=0$. 

\subsection{Exactly solvable model}
\begin{figure}[t]
  \centering
    \includegraphics[width=0.45\textwidth]{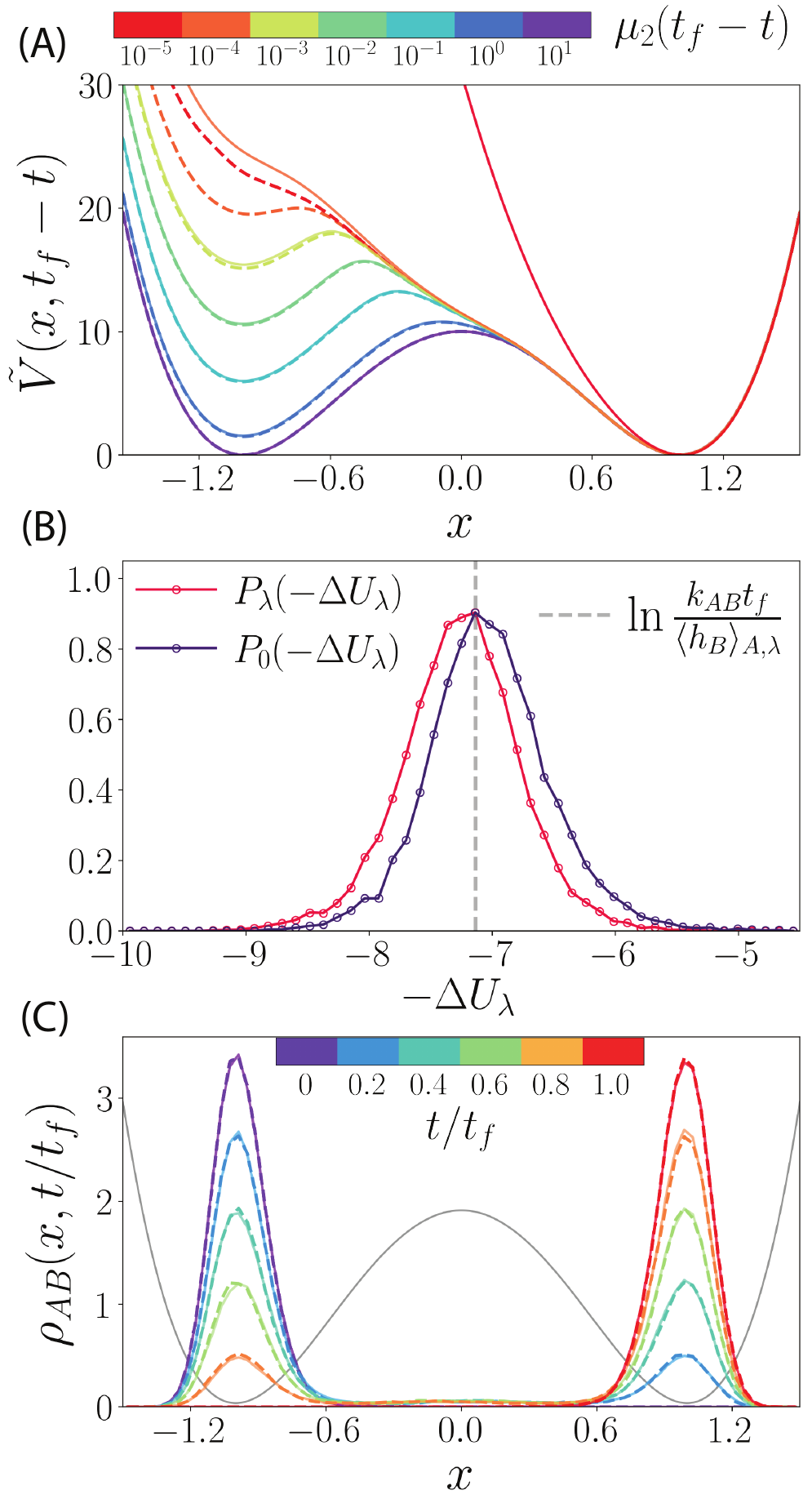}
    \caption{(A) \revisionn{The exact (solid) and the derived (dashed) effective} control potential $\tilde V(x,t_f-t)$ as a function of the position $x$ and the reduced conditioning time $\mu_2 (t_f - t)$.  (B) Action distribution computed within the driven and original reactive ensembles, along with the estimate of $k_{AB}t_f$. (C) Reactive trajectory density for the original reactive dynamics (solid) and the driven reactive dynamics (dashed) as a function of the position and reduced time.}
    \label{fig:1}
\end{figure}

We consider a particle in a bistable double well as an example of an exactly solvable system to study the truncation of the eigenvector expansion and utility of the approximate control force for generating reactive trajectories. We take the potential to be,
\begin{equation}\label{eq:33}
    V(x) = V_0 (x-1)^2(x+1)^2
\end{equation}
where $V_0= 10 \kB T $ and use indicator functions $h_A(x) = \Theta(-x+0.7)$ and $h_B(x) = \Theta(x-0.7)$ to define the two metastable states. Verifying the robustness of the approximate control force for this system is straightforward, since the splitting probability has a closed form,\cite{onsager1938initial}
\begin{align}\label{eq:34}
    \bar q_{B}(x) &= \frac{\int_{x_A}^{x} dx' e^{V(x')/\kbt}}{\int_{x_A}^{x_B} dx' e^{V(x')/\kbt}}\\
    &\approx \frac{1}{2} \left [ 1+\mathrm{erf}(\sqrt{2 V_0/\kB T} x) \right ] \notag
\end{align}
where $x_A=-1$ and $x_B=1$ are the respective minima of the $A$ and $B$ wells. The second line is a harmonic approximation to the barrier, giving an approximate analytic form to the control potential in terms of the error function
\begin{equation}\label{eq:36}
q_{B}(x,t_f-t) \approx \frac{ 1}{2}\left (1+ \mathrm{erf}(\sqrt{2 V_0/\kB T} x) e^{-2 k_{AB}(t_f-t)} \right )
\end{equation}
which is explored in App. \ref{sec:a2}, but in the following we study the numerically exact form using quadrature regularized by truncating and shifting the potential to ensure a continuous force. The rate can also be computed in closed form using Kramer's method\cite{kramers1940brownian,zwanzig2001nonequilibrium} which determines a time dependent effective potential,  \revision{$\tilde V(x,t_f-t) = V(x) - 2 \kbt \ln  q_B(x,t_f-t)$}.

We chose the parameters $\gamma=1.0$, and $\kbt = 1.0$,  which determines a reduced time unit $\tau = \gamma/\kbt$. We use an observation time $t_f=2\tau$ and discretize the dynamics with timestep $dt = 0.001 \tau$. To compute the control force, $\bar q_{B}$ is computed on a grid with $x_A=1.0$,  and fitted using a cubic spline whose derivatives can be taken analytically, $\bar{p}_B = 0.49$ and $\mu_2 = 0.0007173 \tau^{-1}$. Using these parameters, the effective potential $\tilde V(x,t_f-t)$  for different values of the conditioning time $t_f - t$  under the action of the approximate control drift is shown in Fig. \ref{fig:1} (A). When $t_f - t$ is on the order of magnitude of the $1/\mu_2$, the effective potential is simply given by the original potential of the system, since the system is guaranteed to react within those timescales and does not require external control forces to affect the conditioning. As the observation time decreases, the potential within the reactant state is lifted with an exponential dependence making the product well more favorable. Finally at $t=t_f$, the potential becomes unstable within the original reactant well expelling the system out of it. The exact form of the time dependent committor is also shown and becomes harmonic within $B$  and diverges everywhere out of it, ensuring transfer to the product state. The approximate form only guarantees transfer out of $A$ due to the truncation of the eigenvector expansion.

The accuracy of the control forces is assessed by collecting 10000 reactive trajectories generated using brute-force dynamics and 10000 trajectories driven with the control force. For the chosen indicator function, we find that approximately 92\% of the trajectories are reactive. We compute $\Delta U_\lambda$ within the original and the driven reactive trajectory ensemble and consider the distribution of $\Delta U_\lambda$ in both ensembles. This is plotted in  Fig. \ref{fig:1} (B). The two distributions show significant overlap, suggesting  that the two ensembles, the undriven reactive path ensemble and the driven one, are \revision{nearly} statistically indistinguishable. 
The distributions are related through\cite{das2022direct}
\begin{equation}
\frac{ P_\lambda(-\Delta U_{\lambda})}{ P_0(-\Delta U_{\lambda})} =e^{-\Delta U_\lambda+\ln k_{AB} t_f/\langle h_B \rangle_{A,\lambda}}
\end{equation}
which requires that they cross where $\Delta U_\lambda = \ln k_{AB} t_f/\langle h_B \rangle_{A,\lambda} =-7.12 \pm 0.02$. 
The two estimators that we use to determine the accuracy of the force is the first cumulant of $\Delta U_\lambda$. The driven estimator is found to be $-\langle \Delta U_{\lambda} \rangle_{B|A,\lambda} - \langle h_{B}\rangle_{A,\lambda}  = -7.34 \pm 0.01$ and the undriven estimator is computed to be $-\langle \Delta U_{\lambda} \rangle_{B|A,0} - \langle h_{B}\rangle_{A,\lambda}= -7.10 \pm 0.01$, both of which are approximately $0.12$ away from the numerical estimate of $\ln k_{AB}t_f = -7.21 \pm 0.01$. The small displacement is due to the finite timestep, not the approximate representation of the force as explored in App. \ref{sec:a3}, where we compare the action distribution for the approximate form to the exact time-dependent committor computed using two different methods. Notably, we find that both the approximate and the exact control forces incur the same errors in both estimators, and this error can be mitigated to $<$ 0.01 by choosing a smaller timestep. The agreement between the two ensembles can be seen in Fig.  \ref{fig:1} (C), where we plot the reactive trajectory density $\rho_{AB}(x,t/t_f)$ for both the driven and the original dynamics obtained by histogramming the 10000 trajectories in each ensemble. For all different times, we observe the two densities to agree within the thickness of the lines used to plot them.

These results suggest that $q_B(x,t_f-t)$ expressed in terms of the splitting probability offers a strong solution to the path sampling problem for bistable systems. As we discuss in App. \ref{sec:a3}, the only inaccuracy that arises from truncation at second order for this system is the decrease in the reactivity of the driven ensemble when the observation time $t_f$ is chosen to be close to the relaxation timescale of the system $1/\mu_3$. Under such cases, the assumed separation of timescales breaks down, and the contributions to optimal control force from higher eigenfunctions become important to ensure the conditioning. However, even when $t_f$ is chosen to be $0.5 \tau$, where the transition probability grows nonlinearly with time, 40\% of the driven trajectories are reactive and the variational bound is numerically saturated using the approximate controller. 

Finally, we also consider generalization to systems undergoing motion due to the underdamped Langevin equation App. \ref{sec:a3}, where the optimal controller depends on the velocity. We find that the approximate form is inaccurate for small friction coefficient $\gamma$, but becomes exact at larger values, consistent with the homogenization of the underdamped equation of motion under the high friction limit.\cite{singh2023variational, hartmann2014optimal, pavliotis2016stochastic}

\subsection{Dissociation of a colloidal cluster}
\begin{figure}[t]
  \centering
    \includegraphics[width=0.45\textwidth]{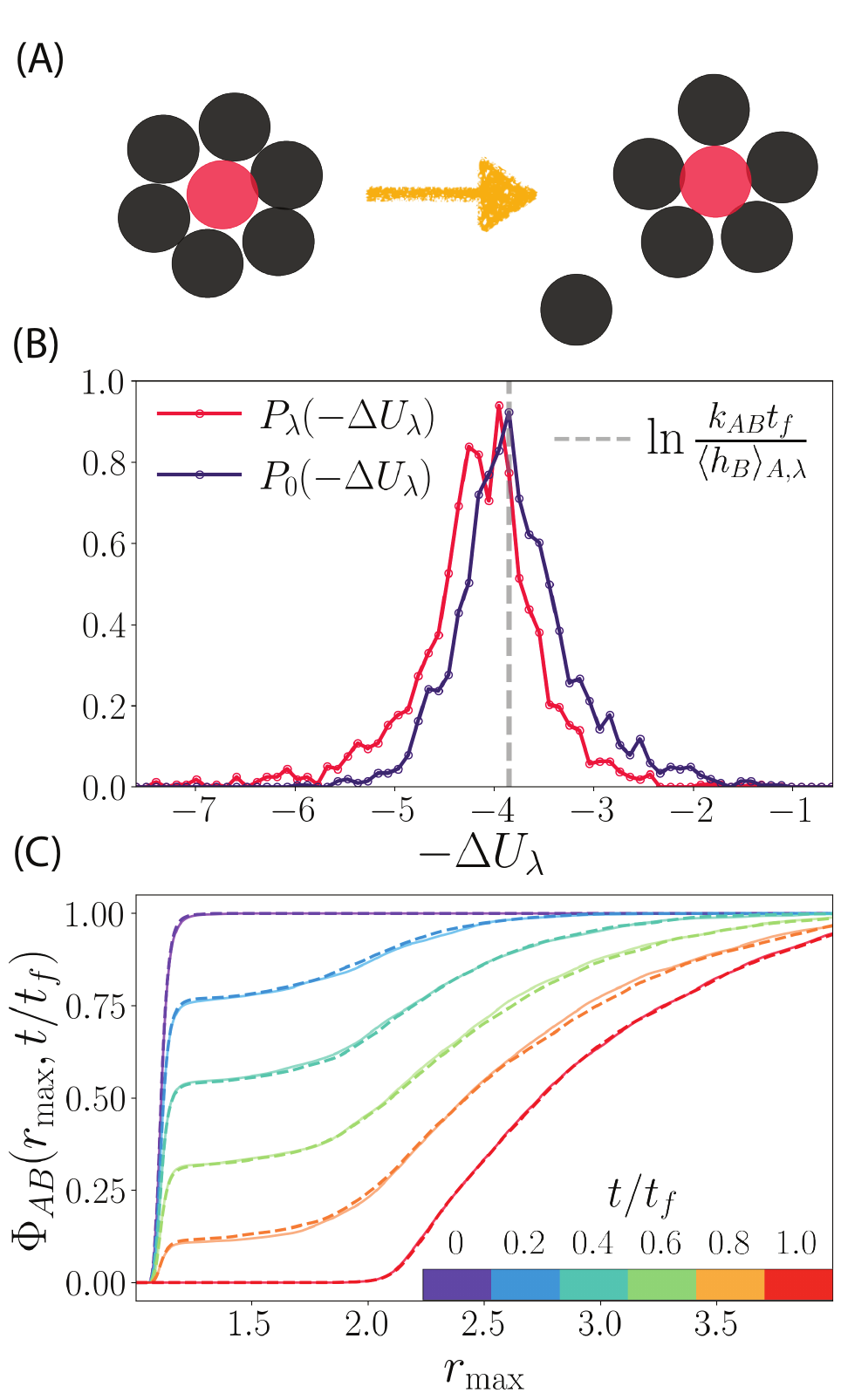}
    \caption{(A) Reactant and produce states for the dissociation of the colloidal cluster. (B) Action distribution computed within the driven and original reactive ensembles, along with the estimate of $k_{AB}t_f$. (C)  Cumulative distribution of the order parameter as a function of time for the original reactive dynamics (solid) and the driven reactive dynamics (dashed).}
    \label{fig:3}
\end{figure}

To study how we can use the the committor to drive reactive transitions in an interacting system where the control force is many-bodied, we consider the dissociation of a colloidal particle from a stable cluster.\cite{das2019variational,das2023nonequilibrium}  Specifically, we consider a model of DNA coated colloids consisting of two components, red and black particles, in two dimensions that form a stable assembly consisting of a particle of one type surrounded by six particles of another type. A total of 7 Brownian particles interact through short ranged repulsions and contact attractions,
\begin{equation}
V(\mathbf{r}) = \frac{1}{2}\sum_{i,j} V_{\mathrm{R}}(r_{ij}) + V_{\mathrm{A}} (r_{ij})
\end{equation}
where $r_{ij}$  is the Euclidian distance between the particles $i$ and $j$. The repulsive potential, $V_{\mathrm{R}}$, is a WCA potential,
\begin{equation}
V_{\mathrm{R}}(r) = 4\epsilon \left [ \left( \frac{\sigma}{r} \right)^{12} -\left(\frac{\sigma}{r} \right)^6 + \frac{1}{4}  \right ] \Theta(2^{1/6}\sigma -r)
\end{equation}
where $\sigma=1$ is the diameter of the particle, $\epsilon$ the energy scale for the repulsion. The attractive potential, $V_{\mathrm{A}}$, is a Morse potential 
\begin{equation}
V_{\mathrm{A}}(r_{ij}) =  D_{ij} (e^{-2a(r_{ij}-r_e)} - 2e^{-a(r_{ij}-r_e)})
\end{equation}
where $D_{ij}/\kbt$ is 24 for unlike particles and 0 for like particles. The particles are placed in square periodic box of size $8\times8\; \sigma^2$. The particles undergo Brownian dynamics with friction coefficient $\gamma = 0.25$ and reduced temperature $\kbt=0.5$. For the potential parameters, we chose $\epsilon/ \kbt = 16 $,  $a \sigma  = 6$ and $r_e=\sigma$. For the system, we chose the reduced time to be $\tau = 2\sigma^2 \gamma /  \kbt = 1$. 

For the parameters that we use the system has two metastable states corresponding to 5 and 6 black particles bound to the red particle, as shown in Fig. \ref{fig:3} (A), and we study the dynamics involved in the $6\rightarrow 5$ reaction. The indicator functions used to defined the $A$ and $B$ basins are $h_A = \Theta(-r_{\mathrm{max}} /\sigma+1.25)$ and $h_B = \Theta(r_{\mathrm{max}}/\sigma -2.10)$, where $r_\mathrm{max}$ is the distance between the red particle and the furthest black particle. This reaction involves a permutation symmetry, which imposes additional constraints on the ansatz used for learning the committor. Further, the dissociated state, where 5 black particles are bound to the red one, is 6-fold degenerate. 

To compute the steady-state committor for the reaction, we expressed it using a neural network ansatz $\bar q_B (\mathbf{r};\boldsymbol{\theta})$ parameterized by $\boldsymbol{\theta}$ and use a variational formalism to optimize the parameters.\cite{khoo2019solving,rotskoff2022active,hasyim2022supervised} The neural network is featurized using the Behler-Parinello symmetry functions,\cite{behler2007generalized} $\{\mathbf{G}\}$ implemented within the package TorchANI.\cite{gao2020torchani} The loss, $\mathcal{I(\boldsymbol{\theta})}$, for the model $\bar q_B (\mathbf{r}; \boldsymbol{\theta} )$ parameterized by $\boldsymbol{\theta}$ takes the following form as derived from the variational formalism of the BKE with absorbing boundary conditions\cite{rotskoff2022active,khoo2019solving,hasyim2022supervised}
\begin{align}
\mathcal{I(\boldsymbol{\theta})}  &=  \frac{1}{N_{(A\cup B)'}}\sum_{\mathbf{r} \in (A\cup B)'} |\nabla_{\mathbf{r}} \bar{q}_B(\mathbf{r};\boldsymbol{\theta})|^2 \bar{p}(\mathbf{r})  \\
	& + \frac{\xi_A }{N_A}  \sum_{\mathbf{r}\in A}\bar q_B (\mathbf{r};\boldsymbol{\theta})^2 +  \frac{\xi_B}{N_B} \sum_{\mathbf{r}\in B}  (1-\bar q_B(\mathbf{r};\boldsymbol{\theta}))^2\notag
\end{align}
where the first sum runs over all the configurations outside basins $A$ and $B$, and the next two sums run over all the configurations in well $A$ and well $B$ respectively. $N_{(A\cup B)'}$,$ N_A$ and $ N_B$ simply correspond to the total number of configurations in each of domains labelled by the subscripts. The parameters $\xi_X$ denote Lagrange multipliers used to satisfy the boundary conditions within the metastable basins. Details of the symmetry functions, parameters and method of optimization is available in the App. \ref{sec:a4}, and the implementation of the code can be found on Github\cite{gith}.

\revision{The rate $\ln k_{AB} $ for the $6\rightarrow 5$ reaction is found to be approximately $-4.9 \pm 0.1$ indicating that only $1\%$ of the trajectories are reactive in the absence of driving for the chosen $t_f = 0.5\tau$.} The backward reaction is faster, and  $\mu_2$ is estimated to be approximately $0.08933 \tau^{-1}$. After training the committor, we use the form of the driving force given in Eq. \ref{eq:27} to drive 2000 trajectories and compute $\Delta U_\lambda$. Approximately 80\% of the trajectories are observed to be reactive 
using $t_f = 0.5\tau$, which can be increased to 90\% if $t_f = \tau$ is used.  The value of the two variational estimators are  $-\langle \Delta U_\lambda \rangle_{B|A,\lambda} - \langle h_B \rangle_{A,\lambda}=-4.40 \pm 0.06$ and $-\langle \Delta U_\lambda \rangle_{B|A,0} - \langle h_B \rangle_{A,\lambda}=-4.05 \pm 0.05$, with a difference of ~$0.35$.  The distribution of $\Delta U_\lambda$ for the two ensembles is shown in Fig. \ref{fig:3}(B), and overlap significantly, indicating a near optimal control force. The ability to generate statistically indistinguishable reacting trajectories within the driven ensemble is substantiated by computing the cumulative distribution of the order parameter, $r_\mathrm{max}$, and comparing its statistics to those within the native reactive ensemble. The distribution of $r_\mathrm{max}$ is given by marginalizing over the reactive density,
$$
\phi_{AB}(r_\mathrm{max},t)=\int d\mathbf{r} \rho_{AB}(\mathbf{r},t) \delta(r_\mathrm{max}-r_\mathrm{max}(\mathbf{r}))
$$
after which the cumulative distribution $\Phi_{AB}(r_\mathrm{max},t)$ is
\begin{equation}\label{eq:crmax}
\Phi_{AB}(r_\mathrm{max},t) = \int_0^{r_\mathrm{max}} dr_\mathrm{max}' \phi_{AB}(r_\mathrm{max}',t) 
\end{equation}
and plotted in Fig.~\ref{fig:3}(C) for $\rho_{AB}$ generated with and without driving. These are visually indistinguishable.

\begin{figure}[t]
  \centering
    \includegraphics[width=0.46\textwidth]{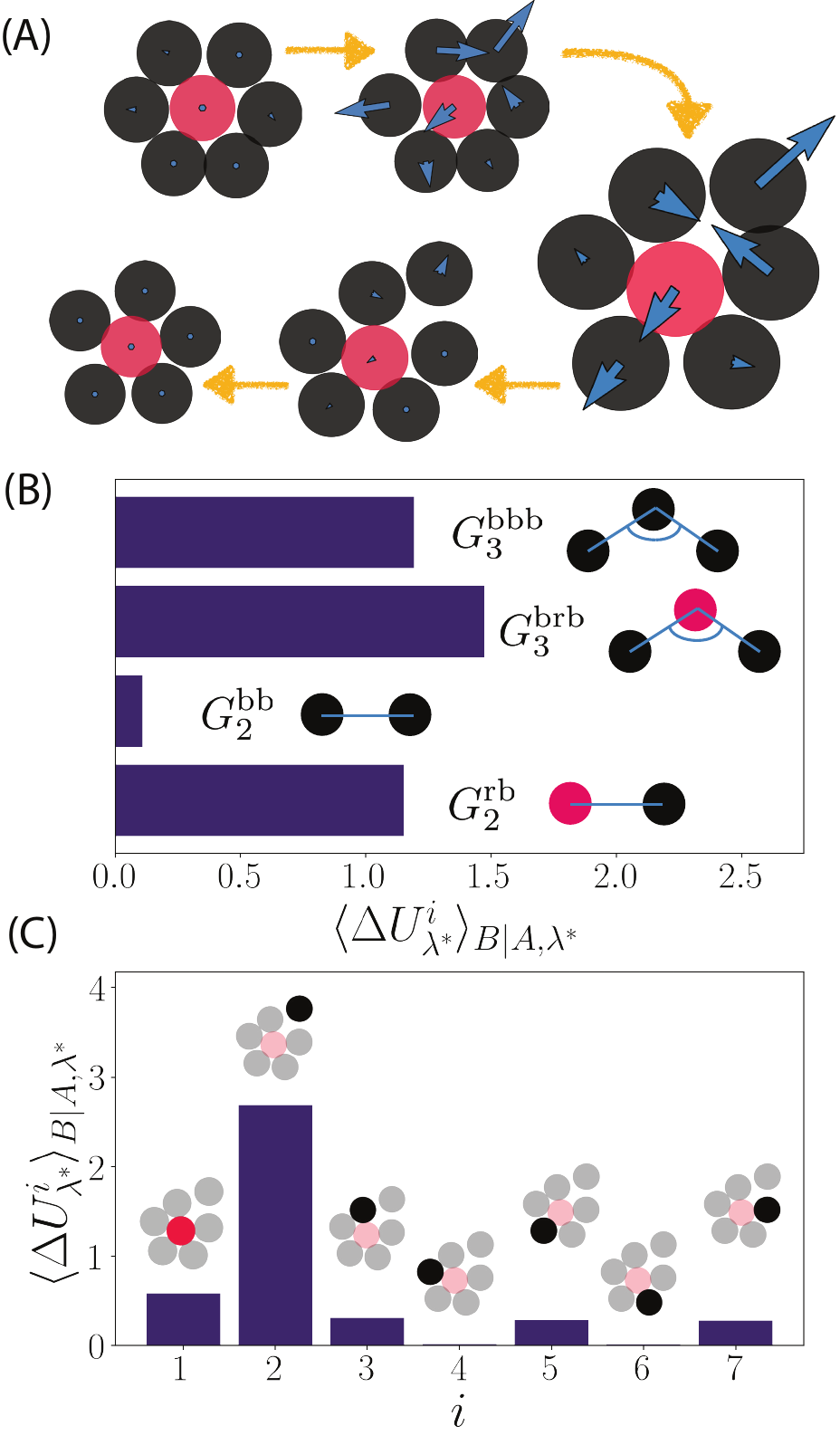}
    \caption{(A) Visualization of the control force given by the rescaled gradients of the splitting probability  for a representative reactive trajectory. Components of the change in action along the set of the symmetry functions (B) and individual particles (C) averaged along the driven reactive trajectory ensemble. }
    \label{fig:4}
\end{figure}

Given the control force nearly saturates the variational bound, we can use it to glean mechanistic information into the reaction. Under the conditioning picture of Doob and Girsanov, these forces encapsulate the natural fluctuations of the system that make a trajectory reactive.\cite{chetrite2015nonequilibrium} A representative trajectory is shown in Fig.~\ref{fig:4}(A), where the control forces are illustrated with arrows with magnitude and direction determined by the gradient of the splitting probability.  As anticipated, we find that the force is many-bodied, acting on multiple particles of the cluster. The basic mechanism of the reaction first involves a natural symmetry breaking by which one black particle undergoes a fluctuation that displaces it from the center red particle. Once this symmetry is broken and a black leaving particle has been determined, forces act between the red and the exiting black particle along the bond vector further separating the two. At the same time, a concerted compressive force arising from the particles closest to the exiting causes them to push the exiting particle away from the cluster. The black particle that is furthest away from the exiting particle contribute a stabilizing force for the center red particle. The direction of the forces show that this importance arises from the fact that the opposite particle lies along the bond vector separating the cluster and exiting particle, and fluctuations of it away from the exiting particle favors reactivity. The mechanism is reminiscent of a T1 transition, a characteristic deformation pattern for switching neighbors in dense media\cite{hasyim2021theory}.

The relative importance of each of these fluctuations, or equivalently the action of the control force, can be quantified using the decomposition in Eq.~\ref{eq:30}, expressed in terms of the symmetry functions used in the neural network or the bare particle indices.  These metrics, shown in Figs. \ref{fig:4} (B) and (C), quantifies the rare fluctuations along each degree of freedom\cite{louwerse2022information} and forms a natural measure to infer the relevance of a descriptor to a reaction.\cite{singh2023variational}  The plot shows that the black-red-black $G_3^{\mathrm{brb}}$ and the black-black-black angular degrees of freedom $G_3^{\mathrm{bbb}}$ are the most important descriptors, followed by the distance symmetry function between the red and black  $G_2^{\mathrm{rb}}$. The black-black $G_2^{\mathrm{bb}}$ function is found be the least important descriptor. We further consider a decomposition of the action along the individual particles and plot it in Fig. \ref{fig:4} (C). The first two particles in Fig. \ref{fig:4} (C) correspond to the red particle and the black particle that is exiting the cluster. The change in action between these two particles accounts for about 70\% of the total action. 


\section{Concluding discussion}

In this work, we have provided an alternate perspective into the role of the committor, its logarithm is a potential that guides the system towards reaction. Similarly, its gradients represent forcefields the application of which guarantees reactivity, ensuring the paths are representative of how the system would evolve naturally, in the absence of an applied force.  Beyond a conceptual illustration, this work addresses multiple problems within importance sampling and rare event simulation.  It offers a strong  solution to the bridge problem in one dimension when the spectral gap assumption holds, since both the leading eigenvectors and eigenvalues can computed exactly using Kramer's method\cite{zwanzig2001nonequilibrium}. This work offers a way to generate uncorrelated reactive trajectories in equilibrium with the same statistics as that of naturally reactive trajectories purely based on configurations obtained from enhanced sampling methods. This is because both the committor and the rate can be computed using the variational form of the boundary value problem.\cite{rotskoff2022active,khoo2019solving,hasyim2022supervised} This work also offers a natural way to infer mechanisms of reactions through the decomposition of the change of action along the system's degrees of freedom, a metric that essentially quantifies the magnitude of fluctuation along a descriptor to guarantee a transition.\cite{singh2023variational} 

Beyond these direct applications, this work also opens up multiple future directions. We believe that this formalism can be easily extended to systems that contain multiple metastable states, and could be used to not only sample reactive trajectories, but also those that occur through specific pathways. It also provides a way to speed up other importance sampling methods, some of which that already perform on-the-fly committor calculations to increase acceptance.\cite{jung2023machine,peters2006obtaining} The formalism employed could be used to better constrain approximate techniques like well-tempered metadynamics,\cite{tiwary2013metadynamics} providing alternative diagnostic measures\cite{salvalaglio2014assessing,khan2020fluxional} of accuracy or rate estimation.

\section*{Data Availability}
The code for the illustration of the splitting probability as the optimal controller for the 1D double well and the DNA labelled colloids along with the training of the neural network ansatz for the latter system can be found on Github. \cite{gith}

\section*{Acknowledgements}
We would like to thank Dr. Avishek Das for invaluable discussions regarding variational path sampling methods and comments regarding the manuscript. This work was supported by the U.S. Department of Energy, Office of Science, Office of Advanced Scientific Computing Research, and Office of Basic Energy Sciences, via the Scientific Discovery through Advanced Computing (SciDAC) program.

\appendix
\setcounter{equation}{0}
\setcounter{figure}{0}
\setcounter{table}{0}
\setcounter{section}{0}

\renewcommand{\theequation}{A\arabic{equation}}
\renewcommand{\thefigure}{A\arabic{figure}}
\renewcommand{\thesection}{A\arabic{section}}

\section{Relationship between $\psi_2^L$ and $\bar q_B$}\label{sec:a1}
\subsection{Constant values of second eigenvector within metastable basins}
The fact that the solution to the boundary value problem can be analytically continued throughout the whole domain is rather subtle and not generally true for arbitrary boundaries. In this section, we use the Feynman-Kac theorem to justify the validity of this approximation when the boundaries are chosen within metastable basins. The arguments presented here exactly follow that of Ryter who first identified the constant values of leading left eigenvectors for metastable systems,\cite{ryter1987eigenfunctions} and similar arguments within the context of chemical physics and applied maths can also be found in multiple references\cite{berezhkovskii2004ensemble,bovier2016metastability,roux2022transition,chen2023discovering}. 

We start by noting that the Feynman-Kac theorem offers a way to estimate the left eigenvectors $\Psi_i^{L}$ of the Backward-Kolmogorov equation by imposing absorbing boundary conditions on some arbitrary domain $D$. If the values of the function on the boundaries of the domain denoted $\partial D$ are known, the values of the function outside of the domain \revisionn{which we denote $D'$} can be computed using
\begin{equation}\label{eq:s1}
    \psi^L_{i}(\mathbf{r} \in D') = \langle \psi_i^L (\mathbf{R}(\tau)) e^{-\mu_i \tau} \rangle_{\mathbf{R}(0)=\mathbf{r}}
\end{equation}
where $\mu_i$ is the $i$th eigenvalue, $\tau$ is the first hitting time on the boundary $\partial D$, and $\langle \cdots \rangle_{\mathbf{R}(0)=\mathbf{r}}$ denotes the expectation value computed over all possible realizations of the corresponding stochastic differential equation $\mathbf{R}$ given the overdamped Langevin equation initiated at $\mathbf{R}(0) = \mathbf{r}$. 

In the case that the domain $D$ consists of 2 disjoint subdomains labelled $D_A$ and $D_B$, the equation can be rewritten as
\begin{align}\label{eq:s2}
    \psi^L_{i}(\mathbf{r} \in D') &= \bar q_A(\mathbf{r}) \langle \psi_i^L (\mathbf{R}(\tau)) e^{-\mu_i \tau_A} \rangle_{\mathbf{R}(0)=\mathbf{r},\tau_A<\tau_B}  \notag\\
    &+ \bar q_B(\mathbf{r}) \langle \psi_i^L (\mathbf{R}(\tau)) e^{-\mu_i \tau_B} \rangle_{\mathbf{R}(0)=\mathbf{r},\tau_B<\tau_A}
\end{align}
where $\bar q_A$ is the conditional probability that the system first hits $D_A$ before hitting $D_B$, or exits out of $D'$ through the subdomain $D_A$, and vice-versa for $\bar q_B$. This function is generally known as the first hitting probability, and can be recognized as the splitting probability when only two absorbing subdomains are considered.

Now, we consider that the system is bistable, and define $R_A$ and $R_B$ as subdomains of the system where it is attracted towards the local minima of the basin with a probability of $1$. The choice of these domains are somewhat arbitrary, and can be intuited in chemical physics by considering how indicator functions are chosen within the Transition Path Sampling formalism.\cite{bolhuis2002transition}

We now consider the form of Equation \ref{eq:s2} when $D_A$ and $D_B$ are placed within regions $R_A$ and $R_B$, such that they form a smaller subset of the region where the system is committed to local minima with a probability of unity. For any $\mathbf{r}\in R_B \cap D_B'$, the probability to first hit the smaller domain $D_B$ rather than $D_A$ is going to be $\approx$ 1, since the system approaches a quasi-stationary state within basin $B$, or in other words,  ergodically samples the basin $B$ before transitioning to $A$. Hence, we have:
\begin{equation}\label{eq:s3}
\revisionn{
    \psi^L_{i}(\mathbf{r} \in R_B \cap D_B') \approx  \langle \psi_i^L (\mathbf{R}(\tau_B)) e^{-\mu_i \tau_B} \rangle_{\mathbf{R}(0)=\mathbf{r}}}
\end{equation}
and a similar form for $\mathbf{r}\in R_A \cap D_A'$. The form above holds for all the left eigenvectors, and can be simplified further for the second left eigenvector by invoking separation of timescales. \revisionn{The timescales of $\tau_B$ for the system initiated within region $R_B \cap D_B'$} correspond to the relaxation timescales of the system within the metastable well $B$, and are multiple orders of magnitude faster than $\mu_2^{-1}$. Hence, $\mu_2 \tau_B \approx 0$, and Eq. \ref{eq:s3} simplifies to:
\begin{equation}\label{eq:s4}
\revisionn{
    \psi^L_{2}(\mathbf{r} \in R_B \cap D_B') \approx  \langle \psi_i^L (\mathbf{R}(\tau_B)) \rangle_{\mathbf{R}(0)=\mathbf{r}}}
\end{equation}
The time-independence of the form above indicates that the values for the second left eigenfunction in region $\psi_2^L(\mathbf{r} \in R_B \cap D_B')$ is constrained to that of the eigenfunction within the boundaries $\psi_2^L(\mathbf{r} \in \partial D_B)$. Now, since $D_B$ is an artificial domain, it can be \revision{constructed such that it encompasses an infinitesimally small sphere. For this case, the value of the eigenfunction for $\mathbf{r} \in \partial D_B$ has be constant with values corresponding to that of the region within $D_B$, since the eigenfunction is continuous and independent of the artificial domain $D_B$.} By Eq. \ref{eq:s4} this also implies that the second eigenvector is constant within the whole subdomain $R_B$ and (through the same arguments) $R_A$ . This form also validates the freedom in choosing the boundaries for solving the boundary value problem when $A$ and $B$ are metastable, since $D_A$ and $D_B$ can be defined arbitrarily as long as they are within $R_A$ and $R_B$.

Denoting $a$ and $b$ as the constant value of $\psi^L_{2}$ within $R_A$ and $R_B$ respectively, and substituting it in Eq. \ref{eq:s2}, we have a solution for the whole domain:
\begin{equation}\label{eq:s5}
    \revisionn{\psi^L_{2}(\mathbf{r} ) \approx a \bar q_A(\mathbf{r})  + b \bar q_B(\mathbf{r}) = a + (b-a) \bar q_B(\mathbf{r})}
\end{equation}
where we have invoked ergodicity of the system $\bar q_B (\mathbf{r}) + \bar q_A(\mathbf{r}) = 1$ to get the final form. A subtlety that we skipped over in going from Eq. \ref{eq:s2} to Eq. \ref{eq:s5} was neglecting the time dependence for regions $R_A'\cup R_B'$. The same separation of timescales argument made in Eq. \ref{eq:s4} can be obtained to neglect time-dependence in the whole domain.
\subsection{Solution for the boundary terms}
To solve for $a$ and $b$, we consider the spectral form for the conditional probability $\langle h_B(\tau) \rangle_A$ where $\mu_3^{-1} < \tau \ll \mu_2^{-1}$
\begin{align}\label{eq:s6}
    \langle h_B(\tau) \rangle_A &= \frac{1}{\bar p_A} \int \int d{\mathbf{r'}} d\mathbf{r} P(\mathbf{r'},\tau|\mathbf{r},0) h_B(\mathbf{r'}) h_A(\mathbf{r}) \bar p (\mathbf{r}) \notag \\
    &\approx \bar p_B + \frac{1}{\bar p_A} \left[ \int d\mathbf{r} \psi_2^L(\mathbf{r'}) h_A(\mathbf{r}) \bar p(\mathbf{r}) \right] \times \notag \\
    & \left[ \int d\mathbf{r'} \psi_2^L(\mathbf{r}) h_B(\mathbf{r'}) \bar p(\mathbf{r'}) \right]  e^{-\mu_2 \tau} \notag \\
    &= \bar p_B \left[1 + ab  e^{-\mu_2 \tau} \right]
\end{align}
where we have used the separation of timescales to truncate the eigenexpansion of the FPE to 2nd order in the second line, and used the fact that $\psi_2^{L}(\mathbf{r})$ is constant within $A$ and $B$ with values $a$ and $b$ to get the final form.
The rate $k_{AB}$ can be computed by taking derivatives of the conditional probability w.r.t. $\tau$ and the form can be simplified by invoking separation of timescale
\begin{equation}\label{eq:s7}
    k_{AB} = \frac{d }{d\tau} \langle h_B(\tau) \rangle_A =  -\bar p_B ab \mu_2  
\end{equation}
The rate can be related to second eigenvalue using the following relationship, $k_{AB}= \bar p_B \mu_2$, which follows from detailed balance $\bar p_A k_{AB} = \bar p_B k_{BA}$ and the bistable assumption $\bar p_A + \bar p_B \approx 1$ in the final step.
Equating the two sides, we have
\begin{equation}\label{eq:s8}
    ab = -1
\end{equation}
A second equation can be constructed by considering the survival probability of the system within B $\langle h_B(\tau) \rangle_B$ under the same separation of timescales for $\tau$
\begin{align}\label{eq:s9}
    \langle h_B(\tau) \rangle_B &= \frac{1}{\bar p_B} \int \int d{\mathbf{r'}} d\mathbf{r} P(\mathbf{r'},\tau|\mathbf{r},0) h_B(\mathbf{r'}) h_B(\mathbf{r}) \bar p (\mathbf{r}) \notag \\
    &\approx \bar p_B \left[ 1+b^2 e^{-\mu_2 \tau} \right]
\end{align}
where we have used the same approximations made in Eq. \ref{eq:s7} to get the final form. Since the system is bistable, the rate of survival is simply related to the rate of the inverse process $k_{BA}$
\begin{equation}\label{eq:s10}
    k_{BA} = - \frac{d }{d\tau} \langle h_B(\tau) \rangle_B =  \bar p_B \mu_2 b^2
\end{equation}
Using the same relation in Eq. \ref{eq:s7} to obtain $k_{BA} = \bar p_A \mu_2$ and equating it with Eq. \ref{eq:s10} to compute $b$, and substituting its value in Eq. \ref{eq:8} to solve for $a$, we have
\begin{equation}\label{eq:s11}
    b = \pm \sqrt{\frac{\bar p_A}{\bar p_B}}, \quad a = \mp \sqrt{\frac{\bar p_B}{\bar p_A}}
\end{equation}
The signs for $a$ and $b$ can be chosen arbitrarily, as long as they are opposite. Choosing $b$ to be positive and substituting into Eq. \ref{eq:s5} and invoking the bistable assumption, we have a closed form for $\psi_2^L(\mathbf{r})$ in terms of $\bar q(\mathbf{r})$,
\begin{align}\label{eq:s12}
    \psi_2^L(\mathbf{r}) &=    \frac{1}{\sqrt{\bar p_A \bar p_B}} (\bar q(\mathbf{r}) - \bar p_B) 
\end{align}
Finally, substituting this form into the time-dependent committor and simplifying, we have
\begin{align}\label{eq:s13}
    q(\mathbf{r},\tau) &=  \bar p_B + (\bar q(\mathbf{r}) - \bar p_B)e^{-\mu_2 \tau} \notag \\
    &= \bar q_B(\mathbf{r}) e^{-\mu_2 \tau} + \bar p_B(1-e^{-\mu_2 \tau})
\end{align}
which is the expression appearing in the main text.

\section{Harmonic Expansion of the approximate controller}\label{sec:a2}

\begin{figure}[t]
  \centering
    \includegraphics[width=8.5cm]{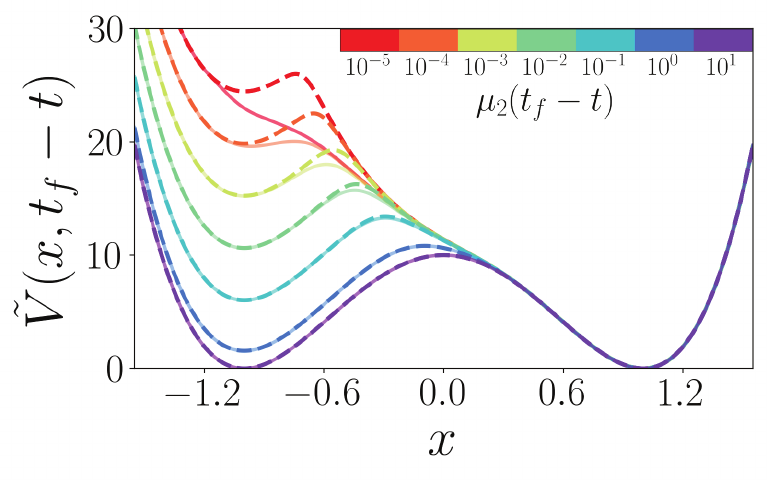}
    \caption{The effective time-dependent potential $\tilde V(x,t_f-t)$ for the approximate controller based on the splitting probability (solid) and the controller constructed using the harmonic approximation for the splitting probability (dashed)}
    \label{fig:D}
\end{figure}

In order to gain physical insight into why the second order truncation of the time-dependent committor is accurate in capturing the natural reaction of the system even though it becomes considerably different than the exact time-dependent committor at short conditioning time, we compare the approximate form to the harmonic expansion of the approximate form given in the main text in Eq. \ref{eq:36}. The plot of the effective potential $\tilde V(x,t_f-t)= V(x) - 2\ln q_B(x,t_f-t)$ for both the approximate and the harmonic form is shown in Fig. \ref{fig:D} (A). The plot shows that the two potentials are accurate at large conditioning time $t_f$, and are generally accurate over the barrier. This is anticipated as the expansion of the harmonic approximation of the committor is performed on the top of the barrier. The only major difference between the potentials is observed at shorter conditioning time where the reactant well is lifted completely for the approximate case such that no barrier is observed between the reactant state and the product state, while the harmonic form still has a small barrier between the two states. This reflects the fact that the harmonic form does not contain any details of the reactant well.

This subtlety in the potential is found to be considerably significant on comparing the estimators of the action between the two forces. Interestingly, only 56\% of the trajectories driven with the harmonic force is found to be reactive, about 36\% lower than the those driven with the approximate form. The values of the estimator is computed to be $-\langle \Delta U_\lambda\rangle_{B|A,\lambda} + \ln \langle h_B\rangle_{A,\lambda} = -7.56$ for the driven case and $-\langle \Delta U_\lambda\rangle_{B|A,0} + \ln \langle h_B\rangle_{A,\lambda} = -6.82$ for the undriven case approximately 0.3-0.4 away from $\ln k_{AB}t_f$, and 0.2-0.3 away for the estimate of the approximate forces. Both the significant decrease in reactivity and the penalty in the change in action in the harmonic form suggests that a crucial part in capturing the time-dependent committor on top of the region near the transition state, is that between the transition state and the reactant state. The destability of the region near the reactant well that increases exponentially as $t \rightarrow t_f$ is found to be the essential part of the time-dependent committor. This is not necessarily surprising, considering that solutions to \revisionn{generalized} bridge problems always exhibit divergences that serves as an attractor towards the region the system is conditioned to end in.\cite{majumdar2015effective} For more complex systems, this result suggests that the region between the reactant and the transition state needs to be fitted accurately. For the system that we considered, we found that using a large Lagrange multiplier for enforcing the boundary values, as well as including some configurations from the reactant and product states to train the variational loss on resolved any issues.

\begin{figure}[b]
  \centering
    \includegraphics[width=8.5cm]{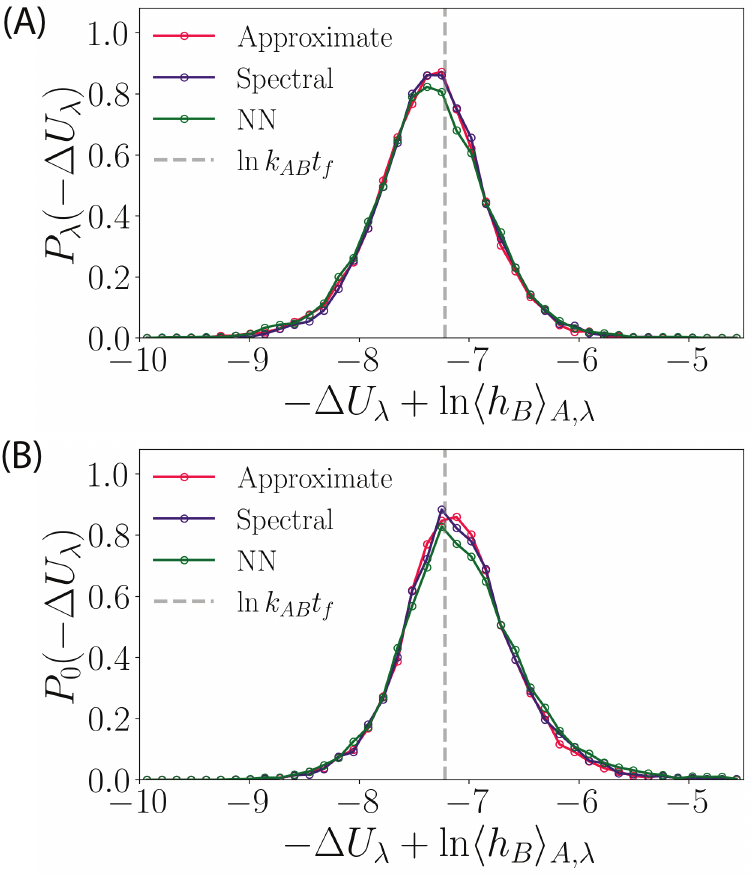}
    \caption{Comparison of the action distribution between the exact control force computed using two different methods and the approximate force formalized in this work. (A) Action distribution with the $\langle h_{B}\rangle_{A,\lambda}$ correction within the driven ensemble and (B) the original ensemble.   }
    \label{fig:A}
\end{figure}

\section{Error analysis for the approximate controller}\label{sec:a3}
We perform an analysis to diagnose the error arising from the second order truncation of the time-dependent committor. Our observation is that the error in the forces arising this truncation only manifests through changes in the reactive probability of the driven reactive trajectory, and leaves the Kullback-Leibler (KL) divergence or the action between the driven reactive trajectory ensemble and the original reactive trajectory ensemble unchanged. 

\subsection{Comparison of the Approximate Controller to the Exact controller}

We first compare the estimates between the approximate controller and the exact controller computed using two different methods. In the first method, we compute the eigenvectors and the eigenvalues spectrally through direct diagonalization of the symmetrized Fokker-Planck operator, and spline fit the forces. For the second method, we prescribe an NN ansatz and train it using variational path sampling formalised within the undriven reactive trajectory ensemble\cite{singh2023variational}. For both of these two cases, 10000  trajectories are obtained by driving the system with the two forces and the statistics from those trajectories are compared to the ones obtained using the approximate form in the main text.

The reactivity of the driven trajectories is found to be 0.995 for the spectral form, 0.975 for the NN ansatz as compared to 0.92 for the approximate form. The distribution of the action within the driven reactive trajectory ensemble with the $\langle h_{B}\rangle_{A,\lambda}$ correction for all the three forces is shown in Fig. \ref{fig:A} (A). The plot shows an excellent overlap for all the three estimators. Notably, the first cumulant of the estimator for all the three methods is computed to be $-7.34$, and hence all the three methods incur an error of ~0.12. Finally the same plot for the probability distribution of the stochastic action within the original reactive trajectory ensemble is shown in Fig. \ref{fig:A} (B), which also shows excellent overlap between the three methods. The first cumulant is computed to be $-7.10$ for all three methods, giving an error of ~$0.12$. This result indicates that the variational bound is saturated numerically, since all three methods incur the same error.

\subsection{Sensitivity to timestep $dt$}
\begin{figure}[t]
  \centering
    \includegraphics[width=8.5cm]{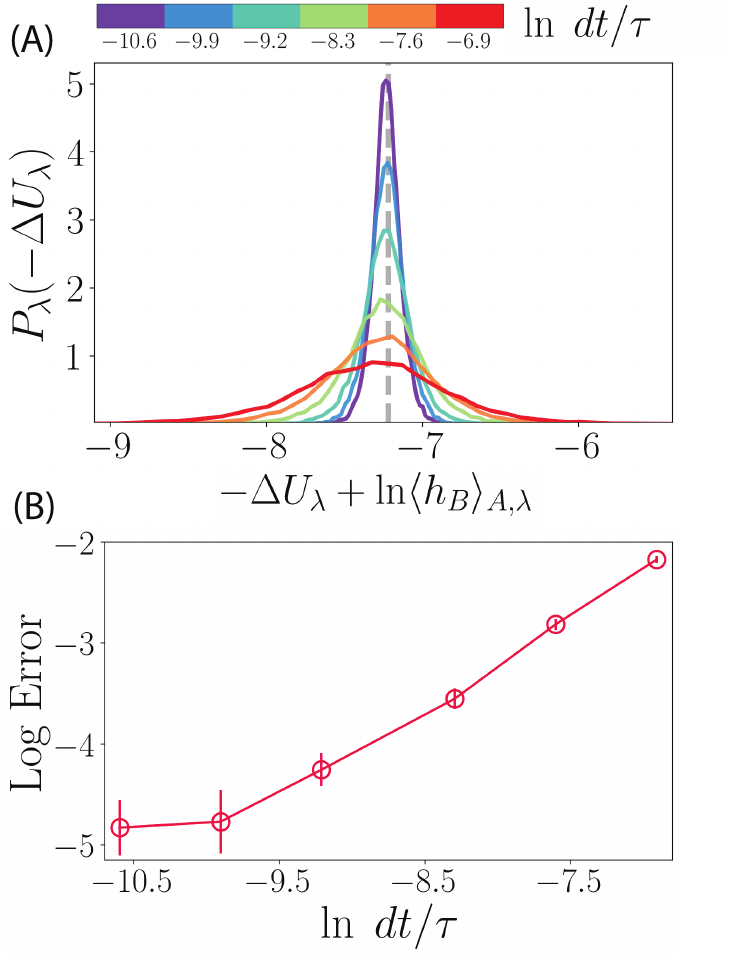}
    \caption{(A) Action distribution computed within the reactive trajectory ensemble driven approximate controller for different timestep discretization $dt$, along with the estimate of $\ln k_{AB}t_f$. (B) The log error in the driven estimator as a function $\ln dt$. }
    \label{fig:C}
\end{figure}

We find that the small error in the driven estimator arises from a finite timestep discretization. To illustrate this, we run 20000 trajectories with $t_f=2\tau$ and a range of $dt$ with the approximate controller. For all values of $dt$, the reactive probability is computed to be the same at approximately 0.92. We first consider the distribution of the difference in stochastic action computed within the driven ensemble and plot it in Fig. \ref{fig:C} (A). We find that the effect of lowering $dt$ culminates in the decrease in variance of the distribution. The variance for the largest $dt$ that corresponding to the one used in the main text is found to be $\approx$ 0.242, which decreases to 0.008 for the smallest $dt$ used. Additionally, we also compare the log difference the between the driven estimate and the log rate $\ln \left( \ln k_{AB}t_f + \langle U_{\lambda} \rangle_{B|A,\lambda} - \langle h_B\rangle_{A,\lambda} \right)$ for all $dt$ and plot it in Fig. \ref{fig:C} (B), along with the standard error computed using bootstrapping. The log error is observed to scale linearly with respect to $\ln dt$, until the second smallest $dt$ considered, after which it is observed to plateau, with a final error estimate of $< 0.01$. As a reference the error for the $dt$ used in the main text was found to be $0.12$.

Both these results can be understood by noting that the saturation of the bound requires the distributions of the change in action within the two ensembles to be a delta function. However, due to finite time discretization, we find the variance of the change in action to scale linearly with $dt$, always leading to a finite width in the distributions. While it is hard to understand the linear scaling of the variance with $dt$, its effect on the error in the estimator given by the first cumulant can be understood easily. The finite width arising from discretization sets a limit to how close to the variational bound one can get, leading to a small but finite correction given by the second order cumulant expansion of the action distributions within the two ensembles. This correction is simply given by the variance of the stochastic action within the two ensembles times a factor of half, which we observe to scale linearly with $dt$. Hence, we conclude the error in estimator for the approximate controller can be attributed solely to timestep error, validating that the approximate controller derived in this work saturates the variational bound, and can be used to obtain a reactive trajectory ensemble that exhibits the exact statistics of naturally reactive trajectories within numerical error.

\subsection{Sensitivity to trajectory length $t_f$}
\begin{figure}[b]
  \centering
    \includegraphics[width=8.5cm]{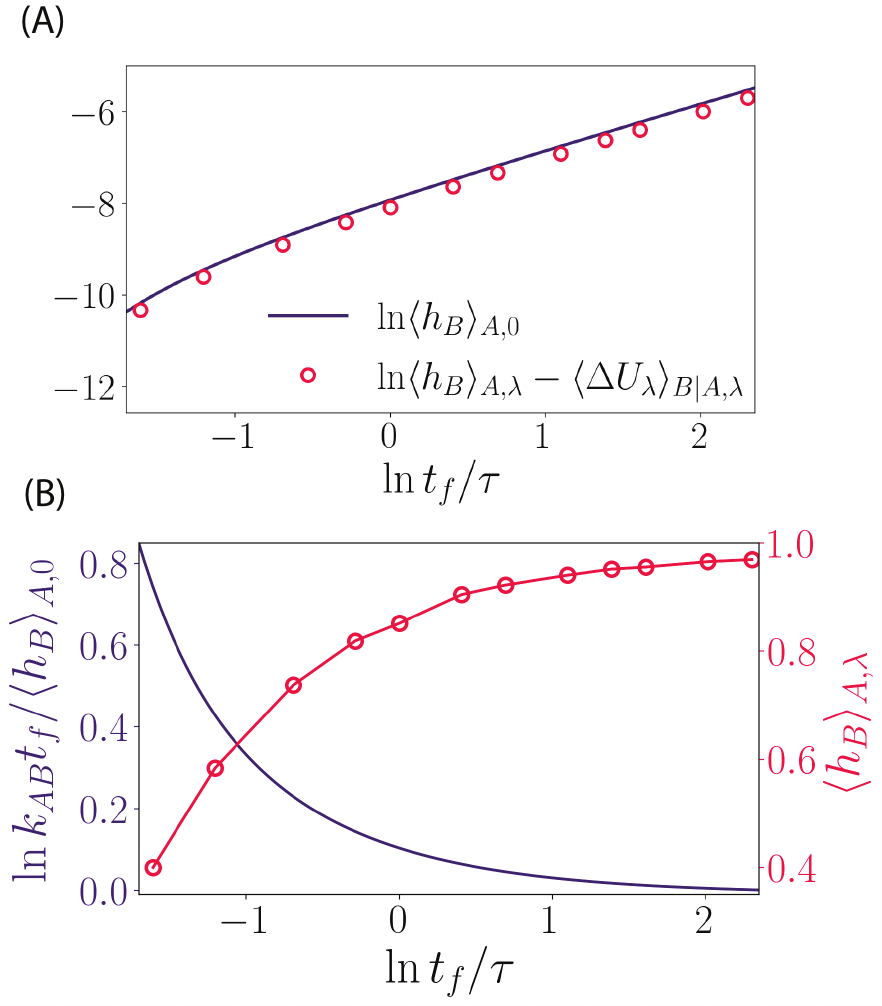}
    \caption{(A) The driven estimator for the approximate controller compared to the original transition probability $\langle h_B \rangle_{A,0}$ as a function of trajectory length $t_f$. (B) Driven reactive probability $\langle h_B \rangle_{A,\lambda}$ along with the log ratio of $k_{AB}t_f$ to the original transition probability $\langle h_B \rangle_{A,0}$ as a function of trajectory length $t_f$. The latter metric quantifies the nonlinear time-dependence of the transition probability when the trajectory length competes with $\mu_3$.}
    \label{fig:B}
\end{figure}

We consider the sensitivity of the approximate controller to the trajectory length $t_f$. The relevant timescale of interest here is given by the time it takes for the conditional reactive probability $\langle h_{B}\rangle_{A,0}$ to enter a linear growth regime where its time-derivative that defines the rate $k_{AB}$ is time-independent. We consider a range of $t_f$ from $0.2 \tau$ to $10 \tau$, and obtain 10000 trajectories by driving the system with the controller expressed in terms of the splitting probability. The driven estimator quantified by the difference in action in the driven reactive trajectory ensemble $\ln \langle h_B \rangle_{A,\lambda}-\langle \Delta U_{\lambda} \rangle_{B|A,\lambda}$ along with the exact conditioned reactive probability $\ln \langle h_{B}\rangle_{A,0}$ is plotted in Fig. \ref{fig:B} (A). The plot shows minimal error between the two estimates that is unaffected by the reaction time, suggesting that the bound is numerically saturated irrespective of the observation time $t_f$ chosen for the reactive trajectory ensemble.

We also consider the reactive probability of the driven process and plot it in Fig. \ref{fig:B} (B). Also plotted is the log ratio of $k_{AB} t_f$ to $\langle h_{B}(t_f)\rangle_{A,0}$, a metric that quantifies the nonlinear time-dependence of the conditioned reactive probability for the chosen $t_f$. For the smallest time $t_f = 0.2 \tau$ the reactivity is found to 40\% which increases quickly to 90\% for $t_f = 0.8 \tau$. For the longest time $t_f=10\tau$ considered, the reactivity is found to be 97\%. Since the indicator function is defined arbitrarily, we also note that for this time, about 99.5\% of the trajectories have final configurations with values $x>0.5$ where the value of $\bar q_B$ is 0.997.

The lower reactive probability can be understood by considering the interplay between $t_f$ and the
truncation of the spectral expansion of the BKE in constructing the controller. On considering the limiting case of $t_f > \mu_2^{-1}$ , the optimal force that makes the system reactive is 0, since the system is guaranteed to transition within those timescales. In other words, under those timescales, the time-dependent committor can be simply truncated to the first order expansion (as given by a constant corresponding to the steady-state distribution of the B well), since the large timescale guarantees reactivity. Now, on considering $t_f \ll \mu_2^{-1}$, driving the trajectories using the first order truncation of the time-dependent committor would give a ratio of reactive trajectories as $k_{AB}t_f$, corresponding to the statistics of the trajectories react naturally. While this ratio is going to be considerably small, the trajectories are going to be indistinguishable from naturally reactive trajectories by construction, since the first order truncation corresponds to no external force.

Now, one can apply the same reasoning to the second order truncation of the time-dependent committor by considering the timescales given by $t_f > \mu_3^{-1}$. This is the timescale in which the second order truncation of the time-dependent committor as expressed in terms of the splitting probability effectively ensures the reactivity of all the trajectories it drives, and more importantly does so in a way that the system reacts naturally. When $t_f$ is chosen such that it competes with $\mu_3$, the ratio of reactive probability goes down, however, the trajectories that do react are going to be indistinguishable from naturally reactive trajectories based on the reasoning made above, and the numerical results shown in this section. In other words, our results suggest that the truncation of the BKE in construction of the optimal controller always saturates the variational bound and penalizes the reactivity of the controlled process.

Finally, we note that even for the extreme case of $t_f=0.2\tau$, where the rate is strongly time-dependent since the log error between $\langle h_{B}(t_f)\rangle_{A,0}$ and $\ln kt_f$ is 0.8, suggesting that the trajectories are instantonic and not representative of the stationary distribution of the reactive trajectory ensemble, the approximate controller increases the reactivity of this specific system by 4 orders of magnitude. This speedup is also arbitrary and would be larger if the $\mu_2$ was lower.

\begin{figure}[t]
  \centering
    \includegraphics[width=0.45\textwidth]{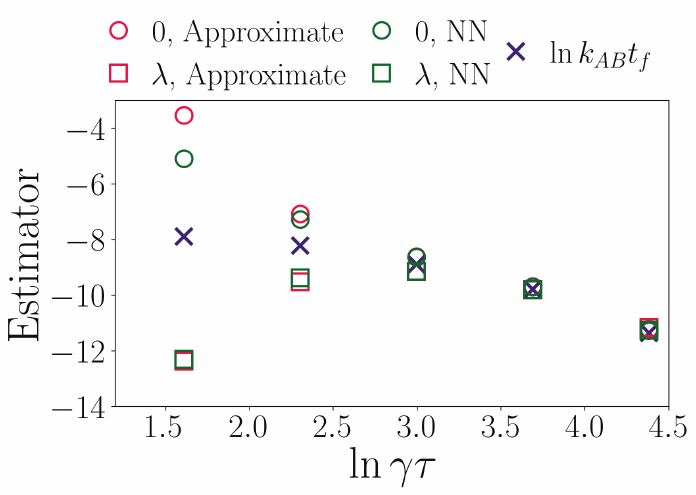}
    \caption{Driven and undriven estimators for the approximate force and the exact velocity independent force, along with the log rate estimate $\ln k_{AB}t_f$ for the underdamped system as a function of $\gamma$. }
    \label{fig:E}
\end{figure}

\subsection{Sensitivity to friction coefficient $\gamma$ for the underdamped Langevin equation}

\setcounter{equation}{13}

\renewcommand{\theequation}{A\arabic{equation}}

We consider application to systems undergoing underdamped dynamics
\begin{equation}\label{eq:under}
m \dot v_i = -\gamma v_i + F_i(\mathbf{r}) + \eta_i \,,\qquad
\dot r_i = v_i
\end{equation}
where $m$ is the mass, $v_i$ is the velocity along the $i$th degree of freedom, $\eta_i$ is gaussian white noise with variance $\langle \eta_i(0) \eta_j(t) \rangle = 2m^{-1}\gamma \kbt \delta_{ij}$ and the rest of the quantities are the same as defined in the main text. While the exact optimal force for underdamped systems is velocity dependent, this dependence can be neglected on taking the underdamped limit of $\gamma^{-1} \to 0$.\cite{hartmann2014optimal,singh2023variational} 
To compare the accuracy of the approximate control force, we consider the same 1D double system as before, but use the underdamped equations of motion given by Eq. \ref{eq:under} with the Normal mode langevin discretization\cite{izaguirre2010multiscale} for a range of friction coefficient $\gamma$. For the parameters, we use $t_f=10\tau, dt=0.01\tau$ and $m=1$.

The rate computed from brute force simulation, along with the value of the undriven and driven estimator for the approximate controller computed from 1000 undriven reactive trajectories and 10000 driven reactive trajectories respectively is shown in Fig. \ref{fig:E} as a function $\gamma$. The plot shows that the error in the estimator decreases systematically with increasing $\gamma$ and vanishes for $\gamma = 20$.  We also consider the estimators for the exact position dependent control forces computed computed using an NN ansatz\cite{singh2023variational} to understand if the error in the estimator is due to the importance of velocities or the truncation of the BKE for constructing the approximate controller. The excellent agreement between both the NN ansatz and the approximate controller indicates that the error in the approximate controller for intermediate $\gamma$ is due to the importance of velocities and not the truncation of the BKE. This result validates the use of the approximate controller for underdamped systems as long as the friction coefficient is large.

\section{Committor optimization for the interacting system}\label{sec:a4}
The variational method for solving the BKE with the boundary value problem is used to solve for the committor\cite{khoo2019solving,rotskoff2022active,hasyim2022supervised}. A neural network ansatz, more specifically the modified version of the Behler-Parinello symmetry functions implemented within the Python package TorchANI are used as descriptors for the model. Two particle types are used for the ansatz, one to represent the red particle and another to represent the black particle. A total of 30 radial symmetry functions are used to represent the distances in the range $[0.9\sigma,4\sqrt2 \sigma]$ with a cutoff of $R_C=4\sqrt2 \sigma$ and gaussian width of $\eta = 16$. A total of $3\times 6$ modified angular symmetry functions are used with 6 angular divisions in the range $[0,2\pi]$, and 3 distance divisions in the range $[0.9\sigma,2\sigma]$ to represent angular information. For the angular symmetry functions $\zeta = 32$, $R_C=2$ and $\eta=8$. All of these parameters are defined in accordance with Equations 3 and 4 of the original ANI paper\cite{smith2017ani}. The structure of the model used is quite similar to the example implementation within the TorchANI documentation, and most of the modifications to the example are made on the final layers to change the objective of the model to learn the committor rather than a potential energy. The number of parameters and descriptors are quite redundant and can be reduced considerably, but that is not explored here.

For learning the committor, the indicator function used to defined the $A$ and $B$ basin is $h_A = \Theta(-r_{\mathrm{max}} /\sigma+1.25)$ and $h_B = \Theta(r_{\mathrm{max}}/\sigma -2.10)$, where $r_\mathrm{max}$ is the distance between the red particle and the furthest black particle. The only modification made within the method is that the region denoted $(A\cup B)'$ on which the variational loss is trained on is defined differently to have slight overlap with samples within the $A$ and $B$ well. Specifically we use the indicator $\Theta(r_{\mathrm{max}} /\sigma - 1.20) \cap \Theta(-r_{\mathrm{max}} /\sigma + 2.2)$ to define region outside the two wells. This modifications has been made to allow the model to learn the location of the boundary automatically. A strong Lagrange multiplier of $\xi_A=\xi_B=5000$ is used for imposing the boundaries. Training was performed for 1000 epochs using 10$^5$ configurations for each domain from a $2.5\times 10^4 \tau$ long trajectory, and the AdamW optimizer\cite{loshchilov2017decoupled} is used for updating the weights and biases. 

\bibliography{SplittingOptimal}

\end{document}